\documentstyle[amssymb,12pt]{article}

\newlength{\minitwocolumn}
\font\teneufm=eufm10
\font\seveneufm=eufm7
\font\fiveeufm=eufm5
\newfam\eufmfam
\textfont\eufmfam=\teneufm
\scriptfont\eufmfam=\seveneufm
\scriptscriptfont\eufmfam=\fiveeufm

\makeatletter
\@addtoreset{equation}{section}
\makeatother

\newtheorem{thm}{Theorem}[section]
\newtheorem{prop}[thm]{Proposition}

\newtheorem{df}{Definition}[section]

\title{\bf
\Large{\bf
Wakimoto realization of Drinfeld current for\\
the
elliptic quantum algebra $U_{q,p}(\widehat{sl}_3)$}
}
\begin{document}
\maketitle
\begin{center}
{
Takeo KOJIMA}
\\~\\
{\it
Department of Mathematics,
College of Science and Technology,
Nihon University,\\
Surugadai, Chiyoda-ku, Tokyo 101-0062, 
JAPAN}
\end{center}
\begin{abstract}
We study a free field realization of the elliptic quantum algebra $U_{q,p}(\widehat{sl_3})$
for arbitrary level $k$.
We give the free field 
realization of elliptic analogue of
Drinfeld current associated with $U_{q,p}(\widehat{sl_3})$
for arbitrary level $k$.
In the limit $p \to 0, q \to 1$ our realization reproduces Wakimoto realization
for the affine Lie algebra $\widehat{sl_3}$.
\end{abstract}
\section{Introduction}

The elliptic quantum group has been proposed in papers \cite{FIJKMY,
Felder, Fronsdal, EF, JKOS1}.
There are two types of elliptic quantum groups,
the vertex type ${\cal A}_{q,p}(\widehat{sl_N})$ and the face type
${\cal B}_{q,\lambda}({g})$, where ${g}$ is a Kac-Moody algebra
associated with a symmetrizable Cartan matrix.
The elliptic quantum groups have the structure of quasi-triangular quasi-Hopf algebras
introduced by V.Drinfeld \cite{Drinfeld}.
H.Konno \cite{Konno} introduced the elliptic quantum algebra $U_{q,p}(\widehat{sl_2})$ 
as an algebra of the screening currents
of the extended deformed Virasoro algebra 
in terms of the fusion SOS model \cite{DJKMO}.
M.Jimbo, H.Konno, S.Odake, J.Shiraishi
\cite{JKOS2} continued to study the elliptic quantum algebra $U_{q,p}(\widehat{sl_2})$.
They constructed the elliptic alnalogue of
Drinfeld currents and identified $U_{q,p}(\widehat{sl_2})$ with
the tensor product of ${\cal B}_{q,\lambda}(\widehat{sl_2})$ and a Heisenberg algebra
${\cal H}$.
The elliptic quantum group ${\cal B}_{q,\lambda}(\widehat{sl_2})$ is a quasi-Hopf algebra
while the elliptic algebra $U_{q,p}(\widehat{sl_2})$ is not.
The intertwining relation of the vertex operator of
${\cal B}_{q,\lambda}(\widehat{sl_2})$ is based on the quasi-Hopf structure of
${\cal B}_{q,\lambda}(\widehat{sl_2})$.
By the above isomorphism $U_{q,p}(\widehat{sl_2})\simeq {\cal B}_{q,\lambda}
(\widehat{sl_2}) \otimes {\cal H}$,
we can understand "intertwining relation" of
the vertex operator for the elliptic algebra $U_{q,p}(\widehat{sl_2})$.
Along the above scheme the elliptic analogue of
Drinfeld current of $U_{q,p}(\widehat{sl_2})$ is extended to those of
$U_{q,p}({g})$ for non-twisted affine Lie algebra ${g}$ \cite{JKOS2, KK}.
In this paper we are interested in higher-rank generalization
of level $k$ free field realization of
the elliptic quantum algebra.
For the elliptic algebra
$U_{q,p}(\widehat{sl_2})$,
there exist two kind of free field realizations
for arbitrary level $k$,
the one is parafermion realization \cite{Konno, JKOS2}, the other is
Wakimoto realization \cite{CD}.
In this paper we are interested in the higher-rank generalization of
Wakimoto realization of $U_{q,p}(\widehat{sl_2})$.
We construct level $k$ free field realization of Drinfeld current
associated with the elliptic algebra $U_{q,p}(\widehat{sl_3})$.
This gives the first example of arbitrary level
free field realization of
the higher-rank elliptic algebra.
This free field realization can be applied for
construction of the integrals of motion
for the elliptic algebra $U_{q,p}(\widehat{sl_3})$.
For this purpose, see references \cite{KS1, KS2, KS3}.

The organization of this paper is as follows.
In section 2 we set the notation and introduce bosons.
In section 3 we review the level $k$ free field realization of the 
quantum group $U_q(\widehat{sl_3})$ \cite{AOS}.
In section 4 we give the level $k$ free field realization of the elliptic 
quantum algebra $U_{q,p}(\widehat{sl_3})$.
In appendix we summarize the normal ordering of the basic operators.

\section{Boson}

The purpose of this section is to set up the basic notation and
to introduce the boson.
In this paper we fix three parameters $q,k,r \in {\mathbb C}$.
Let us set $r^*=r-k$.
We assume $k \neq 0, -3$ and ${\rm Re}(r)>0$, ${\rm Re}(r^*)>0$.
We assume $q$ is a generic with $|q|<1, q \neq 0$.
Let us set a pair of parameters $p$ and $p^*$ by
\begin{eqnarray}
p=q^{2r},~~p^*=q^{2r^*}.\nonumber
\end{eqnarray}
We use the standard symbol of $q$-integer $[n]$ by
\begin{eqnarray}
[n]=\frac{q^n-q^{-n}}{q-q^{-1}}.\nonumber
\end{eqnarray}
Let us set the elliptic theta function $\Theta_p(z)$ by
\begin{eqnarray}
&&\Theta_p(z)=(z;p)_\infty (p/z;p)_\infty (p;p)_\infty,\nonumber\\
&&(z;p)_\infty=\prod_{n=0}^\infty (1-p^nz).\nonumber
\end{eqnarray}
It is convenient to work with the additive notation.
We use the parametrization
\begin{eqnarray}
q&=&e^{-\pi \sqrt{-1}/r \tau},\nonumber\\
p&=&e^{-2\pi \sqrt{-1}/\tau},~~~
p^*=e^{-2\pi \sqrt{-1}/\tau^*},~~(r\tau=r^*\tau^*),\nonumber\\
z&=&q^{2u}.\nonumber
\end{eqnarray}
Let us set Jacobi elliptic theta function $[u]_r, [u]_{r^*}$ by
\begin{eqnarray}
~[u]_r&=&q^{\frac{u^2}{r}-u}\frac{\Theta_{p}(z)}{(p;p)_\infty^3},~~
~[u]_{r^*}=q^{\frac{u^2}{r^*}-u}\frac{\Theta_{p^*}(z)}{(p^*;p^*)_\infty^3}.\nonumber
\end{eqnarray}
The function $[u]_r$ has a zero at $u=0$, enjoys the quasi-periodicity property
\begin{eqnarray}
~[u+r]_r=-[u]_r,~~~~[u+r\tau]_r=-e^{-\pi \sqrt{-1} \tau-
\frac{2\pi \sqrt{-1} u}{r}}[u]_r.\nonumber
\end{eqnarray}
Let us set the delta-function $\delta(z)$ as formal power series.
\begin{eqnarray}
\delta(z)=\sum_{n \in {\mathbb Z}} z^n.\nonumber
\end{eqnarray}
Following \cite{AOS} we introduce free bosons $a_n^1,a_n^2, b_n^1,b_n^2,b_n^3,c_n^1,c_n^2,c_n^3,
(n \in {\mathbb Z}_{\neq 0})$.
\begin{eqnarray}
~[a_n^i,a_m^j]&=&\frac{[(k+3)n][A_{i,j}n]}{n}\delta_{n+m,0},~~
[p_a^i,q_a^j]=(k+3) A_{i,j},~~(i,j=1,2),
\\
~[b_n^i,b_m^j]&=&-\frac{[n]^2}{n} \delta_{i,j}\delta_{n+m,0},~~
[p_b^i,q_b^j]=-\delta_{i,j},~~(i,j=1,2,3),
\\
~[c_n^i,c_m^j]&=&\frac{[n]^2}{n} \delta_{i,j} \delta_{n+m,0},~~
[p_c^i,q_c^j]=\delta_{i,j},~~(i,j=1,2,3).
\end{eqnarray}
Here we have used Cartan matrix
$\left(\begin{array}{cc}
A_{11}&A_{12}\\
A_{21}&A_{22}
\end{array}\right)=\left(\begin{array}{cc}2&-1\\
-1&2
\end{array}\right)$.\\
For parameters $a_1,a_2, b_1,b_2,b_3,c_1,c_2,c_3 \in {\mathbb R}$,
we set
the vacuum vector $|a, b, c\rangle$ of the Fock space 
${\cal F}_{a_1 a_2 b_1 b_2 b_3 c_1 c_2 c_3}$ as following.
\begin{eqnarray}
&&a_n^i|a,b,c\rangle=b_n^j|a,b,c\rangle
=c_n^j|a,b,c\rangle=0,~~(i=1,2; j=1,2,3),
\end{eqnarray}
\begin{eqnarray}
p_a^i|a,b,c\rangle=a_i|a,b,c\rangle,~
p_b^j|a,b,c\rangle=b_j|a,b,c\rangle,~
p_c^j|a,b,c\rangle=c_j|a,b,c\rangle,\nonumber
\\
~~(i=1,2;j=1,2,3;n>0).
\end{eqnarray}
The Fock space
${\cal F}_{a_1 a_2 b_1 b_2 b_3 c_1 c_2 c_3}$ 
is generated by bosons $a_{-n}^1,a_{-n}^2, b_{-n}^1, b_{-n}^2, b_{-n}^3,
c_{-n}^1, c_{-n}^2, c_{-n}^3$ for $n \in {\mathbb N}_{\neq 0}$.
The dual Fock space
${\cal F}_{a_1 a_2 b_1 b_2 b_3 c_1 c_2 c_3}^*$ is defined as the same manner. 
In this paper we construct the elliptic analogue of Drinfeld current
for $U_{q,p}(\widehat{sl_3})$ by these bosons $a_n^i, b_n^j, c_n^j$
acting on the Fock space.

\section{Free Field Realization of $U_q(\widehat{sl_3})$}

The purpose of this section is to give the free
field realization of the quantum affine algebra $U_{q}(\widehat{sl_3})$.
We give a review of Wakimoto realization of $U_q(\widehat{sl_3})$ \cite{AOS}.
Let us set the bosonic operators $a_\pm^i(z), b_\pm^i(z)$,
$\gamma^i(z), \beta_s^i(z)$ by
\begin{eqnarray}
a_\pm^i(z)&=&\pm(q-q^{-1})\sum_{n>0}a_{\pm n}^i z^{\mp n} \pm p_a^i {\rm log}q,~~(i=1,2),
\\
b_\pm^i(z)&=&\pm(q-q^{-1})\sum_{n>0}b_{\pm n}^i z^{\mp n} \pm p_b^i {\rm log}q,~~
(i=1,2,3),
\\
b^i(z)&=&-\sum_{n \neq 0}\frac{b_n^i}{[n]}z^{-n}+q_b^i+p_b^i{\rm log}z,~~(i=1,2,3),
\\
c^i(z)&=&-\sum_{n \neq 0}\frac{c_n^i}{[n]}z^{-n}+q_c^i+p_c^i{\rm log}z,~~(i=1,2,3),
\\
\gamma^i(z)&=&-\sum_{n \neq 0}\frac{(b+c)_n^i}{[n]}z^{-n}
+(q_b^i+q_c^i)+(p_b^i+p_c^i){\rm log}(-z),~~
(i=1,2,3),
\\
\beta_1^i(z)&=&b_+^i(z)-(b^i+c^i)(qz),~\beta_2^i(z)=b_-^i(z)-(b^i+c^i)(q^{-1}z),~~
(i=1,2,3),
\\
\beta_1^i(z)&=&b_+^i(z)+(b^i+c^i)(qz),~\beta_2^i(z)=b_-^i(z)+(b^i+c^i)(q^{-1}z),~~
(i=1,2,3).
\end{eqnarray}
We give a free field realiztaion of Drinfeld current
for $U_q(\widehat{sl_3})$.

\begin{df}~~
We define the bosonic operators $e_1^+(z), e_2^+(z), e_1^-(z), e_2^-(z)$ by
\begin{eqnarray}
e_1^+(z)&=&\frac{-1}{(q-q^{-1})z}(e_1^{+,1}(z)-e_1^{+,2}(z)),
\\
e_2^+(z)&=&\frac{-1}{(q-q^{-1})z}(e_2^{+,1}(z)-e_2^{+,2}(z)+
e_2^{+,3}(z)-e_2^{+,4}(z)),
\\
e_1^-(z)&=&\frac{-1}{(q-q^{-1})z}
(e_1^{-,1}(z)-e_1^{-,2}(z)-e_1^{-,3}(z)+e_1^{-,4}(z)),
\\
e_2^-(z)&=&\frac{-1}{(q-q^{-1})z}(e_2^{-,1}(z)-e_2^{-,2}(z)+e_2^{-,3}(z)-e_2^{-,4}(z)).
\end{eqnarray}
\begin{eqnarray}
\psi^\pm_1(z)&=&:\exp\left(
b_\pm^1(q^{\pm k}z)+b_\pm^1(q^{\pm (k+2)}z)+b_\pm^2(q^{\pm (k+3)}z)-b_\pm^3(q^{\pm (k+2)}z)
+a_\pm^1(q^{\pm \frac{k+3}{2}}z)
\right):,\nonumber\\
\\
\psi^\pm_2(z)&=&:\exp\left(
-b_\pm^1(q^{\pm (k+1)}z)+b_\pm^2(q^{\pm k}z)+b_\pm^3(q^{\pm (k+1)}z)
+b_\pm^3(q^{\pm (k+3)}z)
+a_\pm^2(q^{\pm \frac{k+3}{2}}z)
\right):,\nonumber\\
\end{eqnarray}
Here we have set
\begin{eqnarray}
e_1^{+,1}(z)&=&:\exp\left(\beta_1^1(z)\right):,\\
e_1^{+,2}(z)&=&:\exp\left(\beta_2^1(z)\right):,\\
e_2^{+,1}(z)&=&:\exp\left(\gamma^1(z)+\beta_1^2(z)\right):,\\
e_2^{+,2}(z)&=&:\exp\left(\gamma^1(z)+\beta_2^2(z)\right):,\\
e_2^{+,3}(z)&=&:\exp\left(\beta_1^3(qz)+b_+^2(z)-b_+^1(qz)\right):,\\
e_2^{+,4}(z)&=&:\exp\left(\beta_2^3(qz)+b_+^2(z)-b_+^1(qz)\right):,\\
\nonumber
\\
e_1^{-,1}(z)&=&:\exp\left(\beta_4^1(q^{-k-2}z)+b_-^2(q^{-k-3}z)-
b_-^3(q^{-k-2}z)+
a_-^1(q^{-\frac{k+3}{2}}z)
\right):,
\\
e_1^{-,2}(z)&=&:\exp\left(
\beta_3^1(q^{k+2}z)+b_+^2(q^{k+3}z)-b_+^3(q^{k+2}z)+a_+^1(q^{\frac{k+3}{2}}z)
\right):,
\\
e_1^{-,3}(z)&=&
:\exp\left(
\gamma^2(q^{k+2}z)
+\beta_1^3(q^{k+2}z)
+b_+^2(q^{k+3}z)-b_+^3(q^{k+2}z)
+a_+^1(q^{\frac{k+3}{2}}z
\right):,\\
e_1^{-,4}(z)&=&
:\exp\left(
\gamma^2(q^{k+2}z)
+\beta_2^3(q^{k+2}z)
+b_+^2(q^{k+3}z)-b_+^3(q^{k+2}z)
+a_+^1(q^{\frac{k+3}{2}}z
\right):,
\\
e_2^{-,1}(z)&=&
:\exp\left(
\gamma^2(q^{-k-1}z)-\beta_3^1(q^{-k-1}z)
+2b_-^3(q^{-k-1}z)+a_-^2(q^{-\frac{k+3}{2}}z)
\right):,
\\
e_2^{-,2}(z)&=&
:\exp\left(
\gamma^2(q^{-k-1}z)-\beta_4^1(q^{-k-1}z)
+2b_-^3(q^{-k-1}z)+a_-^2(q^{-\frac{k+3}{2}}z)
\right):,
\\
e_2^{-,3}(z)&=&:\exp\left(\beta_4^3(q^{-k-3}z)+a_-^2(q^{-\frac{k+3}{2}}z)
\right):,
\\
e_2^{-,4}(z)&=&:\exp\left(\beta_3^3(q^{k+3}z)+a_+^2(q^{\frac{k+3}{2}}z)\right):.
\end{eqnarray}
\end{df}
Here the symbol $:{\cal O}:$ represents the normal ordering
of ${\cal O}$.
For example we have
\begin{eqnarray}
:b_k b_l:=\left\{
\begin{array}{cc}
b_k^i b_l^i,& k<0\\
b_l^i b_k^i,& k>0.
\end{array}\right.~~~
:p_b^i q_b^i:=:q_b^i p_b^i:=q_b^i p_b^i.\nonumber
\end{eqnarray}

\begin{thm}~~\cite{AOS}~~
The bosonic operators $e_i^\pm(z)$,
$\psi_i^\pm(z)$, $(i=1,2)$
satisfy the following commutation relations.
\begin{eqnarray}
(z_1-q^{A_{i,j}}z_2)e_i^+(z_1)e_j^+(z_2)&=&(q^{A_{i,j}}z_1-z_2)
e_j^+(z_2)e_i^+(z_1),\\
(z_1-q^{-A_{i,j}}z_2)e_i^-(z_1)e_j^-(z_2)&=&(q^{-A_{i,j}}z_1-z_2)
e_j^-(z_2)e_i^-(z_1),\\
~[\psi_i^\pm(z_1),\psi_j^\pm(z_2)]&=&0,
\end{eqnarray}
\begin{eqnarray}
&&(z_1-q^{A_{i,j}-k}z_2)(z_1-q^{-A_{i,j}+k}z_2)\psi_i^\pm(z_1)\psi_j^\mp(z_2)\nonumber\\
&=&(z_1-q^{A_{i,j}+k}z_2)(z_1-q^{-A_{i,j}-k}z_2)\psi_j^\mp(z_2)\psi_i^\pm(z_1),
\end{eqnarray}
\begin{eqnarray}
(z_1-q^{\pm (A_{i,j}-\frac{k}{2})}z_2)\psi_i^+(z_1)e^\pm_j(z_2)&=&
(q^{\pm A_{i,j}}z_1-q^{\mp \frac{k}{2}}z_2)e^\pm_j(z_2)\psi_i^+(z_1),\\
(z_1-q^{\pm (A_{i,j}-\frac{k}{2})}z_2)e^\pm_i(z_1)\psi_j^-(z_2)&=&
(q^{\pm A_{i,j}}z_1-q^{\mp \frac{k}{2}}z_2)\psi_j^-(z_2)e^\pm_i(z_1),
\end{eqnarray}
\begin{eqnarray}
&&\left\{
e_i^\pm(z_1)e_i^\pm(z_2)e_j^\pm(z_3)-(q+q^{-1})
e_i^\pm(z_1)e_j^\pm(z_3)e_j^\pm(z_2)+
e_i^\pm(z_3)e_i^\pm(z_1)e_j^\pm(z_2)
\right\}\nonumber\\
&&+\left\{ z_1 \leftrightarrow z_2 \right\}=0,~~{\rm for}~~(i \neq j),
\end{eqnarray}
\begin{eqnarray}
&&[e_i^+(z_1),e_j^-(z_2)]=\frac{\delta_{i,j}}{(q-q^{-1})z_1 z_2}
\left(\delta\left(q^{-k}\frac{z_1}{z_2}\right)\psi_i^+(q^{-\frac{k}{2}}z_1)
-\delta\left(q^k\frac{z_1}{z_2}\right)
\psi_i^-(q^{-\frac{k}{2}}z_2)\right).\nonumber\\
\end{eqnarray}
\end{thm}
Hence $e_i^\pm(z), \psi_i^\pm(z)$ give level $k$ free field realization of 
$U_q(\widehat{sl_3})$.

\section{Free Field Realization of $U_{q,p}(\widehat{sl_3})$}

The purpose of this section is to give
a free field realization of the elliptic analogue of
Drinfeld current for $U_{q,p}(\widehat{sl_3})$ with arbitrary level $k\neq 0,-3$.
Let us set the bosonic operators 
${\cal B}_\pm^{* i}(z), {\cal B}_\pm^{i}(z), (i=1,2,3)$,
${\cal A}^{* i}(z), {\cal A}^{i}(z), (i=1,2)$ by
\begin{eqnarray}
{\cal B}_\pm^{* i}(z)&=&\exp\left(\pm \sum_{n>0} \frac{b_{-n}^i}{[r^*n]}z^n\right),
~~(i=1,2,3),
\\
{\cal B}_\pm^{ i}(z)&=&\exp\left(\pm \sum_{n>0} \frac{b_n^i}{[rn]} z^{-n}\right),
~~(i=1,2,3),
\\
{\cal A}^{i *}(z)&=&\exp\left(\sum_{n>0}\frac{a_{-n}^i}{[r^*n]}z^{n}\right),
~~(i=1,2),
\\
{\cal A}^i(z)&=&\exp\left(-\sum_{n>0}\frac{a_n^i}{[rn]}z^{-n}\right),~~(i=1,2).
\end{eqnarray}

\begin{df}~~Let us set the bosonic operators $e_i(z), f_i(z), \Psi_i^\pm(z),
(i=1,2)$ by
\begin{eqnarray}
&&e_i(z)={U}^{* i}(z)e_i^+(z),~~(i=1,2),\\
&&f_i(z)=e_i^-(z){U}^i(z),~~(i=1,2),\\
&&\Psi_i^+(z)= U^{*i}(q^{\frac{k}{2}}z)
\psi_i^+(z)U^i(q^{-\frac{k}{2}}z), ~~(i=1,2),\\
&&\Psi_i^+(z)=U^{*i}(q^{-\frac{k}{2}}z)\psi_i^-(z)
U^i(q^{\frac{k}{2}}z),~~(i=1,2).
\end{eqnarray}
Here we have set
\begin{eqnarray}
{U}^{* 1}(z)&=&
{\cal B}_+^{* 1}(q^{r^*}z){\cal B}_+^{* 1}(q^{r^*-2}z)
{\cal B}_+^{* 2}(q^{r^*-3}z)
{\cal B}_-^{* 3}(q^{r^*-2}z){\cal A}^{* 1}(q^{r^*+\frac{k-3}{2}}z),
\\
{U}^{* 2}(z)&=&
{\cal B}_+^{* 3}(q^{r^*-3}z){\cal B}_+^{* 3}(q^{r^*-1}z)
{\cal B}_+^{* 2}(q^{r^*}z)
{\cal B}_-^{* 1}(q^{r^*-1}z){\cal A}^{* 2}(q^{r^*+\frac{k-3}{2}}z),
\\
{U}^1(z)&=&{\cal B}_-^{1}(q^{-r^*}z){\cal B}_-^1(q^{-r^*+2}z)
{\cal B}_-^2(q^{-r^*+3}z){\cal B}_+^3(q^{-r^*+2}z)
{\cal A}^{1}(q^{-r^*-\frac{k-3}{2}}z),
\\
{U}^2(z)&=&{\cal B}_-^3(q^{-r^*+1}z){\cal B}_-^3(q^{-r^*+1}z)
{\cal B}_-^2(q^{-r^*}z){\cal B}_+^1(q^{-r^*+1}z)
{\cal A}^2(q^{-r^*-\frac{k-3}{2}}z).
\end{eqnarray}
\end{df}
The above free field realization
of the twistors $U^{*i}(z), U^i(z)$, $(i=1,2)$ is the 
main result of this paper.

\begin{prop}~~The bosonic operators $e_i(z), f_i(z), \Psi_i^\pm(z)$, 
$(i=1,2)$ satisfy
the following commutation relations.
\begin{eqnarray}
e_i(z_1)e_j(z_2)&=&
q^{-A_{i,j}}\frac{\Theta_{p^*}(q^{A_{i,j}}z_1/z_2)}
{\Theta_{p^*}(q^{-A_{i,j}}z_1/z_2)}
e_j(z_2)e_i(z_1),
\\
f_i(z_1)f_j(z_2)&=&
q^{A_{i,j}}\frac{\Theta_{p}(q^{-A_{i,j}}z_1/z_2)}{\Theta_{p}(q^{A_{i,j}}z_1/z_2)}
f_j(z_2)f_i(z_1),
\end{eqnarray}
\begin{eqnarray}
\Psi_i^\pm(z_1)\Psi_j^\pm(z_2)&=&\frac{\Theta_p(q^{-A_{i,j}}z_1/z_2)
\Theta_{p^*}(q^{A_{i,j}}z_1/z_2)}{
\Theta_p(q^{A_{i,j}}z_1/z_2)\Theta_{p^*}(q^{-A_{i,j}}z_1/z_2)}
\Psi_j^\pm(z_2)\Psi_i^\pm(z_1),\\
\Psi_i^\pm(z_1)\Psi_j^\mp(z_2)&=&
\frac{
\Theta_p(pq^{-A_{i,j}-k}z_1/z_2)
\Theta_{p^*}(p^*q^{A_{i,j}+k}z_1/z_2)
}{
\Theta_p(pq^{A_{i,j}-k}z_1/z_2)
\Theta_{p^*}(p^*q^{-A_{i,j}+k}z_1/z_2)
}
\Psi_j^\mp(z_2)\Psi_i^\pm(z_1),
\\
\nonumber
\\
\Psi_i^\pm(z_1)e_j(z_2)&=&
\frac{\Theta_{p^*}(q^{A_{i,j}\pm \frac{k}{2}}z_1/z_2)}{
\Theta_{p^*}(q^{-A_{i,j}\pm \frac{k}{2}}z_1/z_2)}
e_j(z_2)\Psi_i^\pm(z_1),\\
\Psi_i^\pm(z_1)f_j(z_2)&=&
\frac{\Theta_{p^*}(q^{-A_{i,j}\mp \frac{k}{2}}z_1/z_2)}{
\Theta_{p^*}(q^{A_{i,j}\mp \frac{k}{2}}z_1/z_2)}
e_j(z_2)\Psi_i^\pm(z_1),
\end{eqnarray}
\begin{eqnarray}
~[e_i(z_1),f_j(z_2)]=\frac{\delta_{i,j}}
{(q-q^{-1})z_1 z_2}\left(
\delta\left(q^{-k}\frac{z_1}{z_2}\right)
\Psi_i^+(q^{-k/2}z_1)-
\delta\left(q^{k}\frac{z_1}{z_2}\right)
\Psi_i^-(q^{-k/2}z_2)\right),\nonumber\\
(i\neq j).~~~~
\end{eqnarray}
\end{prop}
We introduce the Heisenberg algebra ${\cal H}$ generated by the following
$P_i,Q_i$, $(i=1,2)$.
\begin{eqnarray}
~[P_i,Q_j]=\frac{A_{i,j}}{2},~~(i,j=1,2).
\end{eqnarray}
\begin{df}~~Let us define the bosonic operators
$E_i(z), F_i(z), H_i^\pm(z)
\in U_q(\widehat{sl_3}){\otimes}{\cal H}$, $(i=1,2)$ by
\begin{eqnarray}
E_1(z)&=&e_1(z)e^{2Q_1}z^{-\frac{P_1-1}{r^*}},~~
E_2(z)=e_2(z)e^{2Q_2}z^{-\frac{P_2-1}{r^*}},\\
F_1(z)&=&f_1(z)z^{\frac{2p_b^1+p_b^2-p_b^3+p_a^1}{r}}z^{\frac{P_1-1}{r}},~~
F_2(z)=f_2(z)z^{\frac{2p_b^3+p_b^2-p_b^1+p_a^2}{r}}z^{\frac{P_2-1}{r}},
\\
H_1^\pm(z)&=&\Psi_1^\pm(z)e^{2Q_1}
(q^{\mp \frac{k}{2}}z)^{\frac{2p_b^1+p_b^2-p_b^3+p_a^1}{r}}
(q^{\pm (r-\frac{k}{2})}z)^{\frac{P_1-1}{r}-\frac{P_1-1}{r^*}},\\
H_2^\pm(z)&=&\Psi_2^\pm(z)e^{2Q_2}
(q^{\mp \frac{k}{2}}z)^{\frac{2p_b^3+p_b^2-p_b^1+p_a^2}{r}}
(q^{\pm (r-\frac{k}{2})}z)^{\frac{P_2-1}{r}-\frac{P_2-1}{r^*}}.
\end{eqnarray}
\end{df}

\begin{thm}~~The bosonic operators $E_i(z), F_i(z), H_i^\pm(z)$, $(i=1,2)$
satisfy the following commutation relations.
\begin{eqnarray}
E_i(z_1)E_j(z_2)&=&\frac{\displaystyle
\left[u_1-u_2+\frac{A_{i,j}}{2}\right]_{r^*}}
{\displaystyle
\left[u_1-u_2-\frac{A_{i,j}}{2}\right]_{r^*}}E_j(z_2)E_i(z_1),
\\
F_i(z_1)F_j(z_2)&=&
\frac{\displaystyle
\left[u_1-u_2-\frac{A_{i,j}}{2}\right]_{r}}
{\displaystyle
\left[u_1-u_2+\frac{A_{i,j}}{2}\right]_{r}}F_j(z_2)F_i(z_1),\\
H^\pm_i(z_1)H^\pm_j(z_2)&=&\frac{\displaystyle
\left[u_1-u_2-\frac{A_{i,j}}{2}\right]_r
\left[u_1-u_2+\frac{A_{i,j}}{2}\right]_{r^*}}{
\displaystyle
\left[u_1-u_2+\frac{A_{i,j}}{2}\right]_r
\left[u_1-u_2-\frac{A_{i,j}}{2}\right]_{r^*}}
H^\pm_j(z_2)H^\pm_i(z_1),\\
H^+_i(z_1)H^-_j(z_2)&=&
\frac{\displaystyle
\left[u_1-u_2-\frac{A_{i,j}}{2}-\frac{k}{2}\right]_r
\left[u_1-u_2+\frac{A_{i,j}}{2}+\frac{k}{2}\right]_{r^*}}{
\displaystyle
\left[u_1-u_2+\frac{A_{i,j}}{2}-\frac{k}{2}\right]_r
\left[u_1-u_2-\frac{A_{i,j}}{2}+\frac{k}{2}\right]_{r^*}}
H^-_j(z_2)H^+_i(z_1),\nonumber\\
\\
H^\pm_i(z_1)E_j(z_2)&=&
\frac{\displaystyle
\left[u_1-u_2\pm\frac{k}{4}+\frac{A_{i,j}}{2}\right]_{r^*}
}
{\displaystyle
\left[u_1-u_2\pm \frac{k}{4}-\frac{A_{i,j}}{2}\right]_{r^*}}
E_j(z_2)H^\pm_i(z_1),\\
H^\pm_i(z_1)F_j(z_2)&=&
\frac{\displaystyle
\left[u_1-u_2\mp\frac{k}{4}-\frac{A_{i,j}}{2}\right]_{r}
}
{
\displaystyle
\left[u_1-u_2\mp \frac{k}{4}+\frac{A_{i,j}}{2}\right]_{r}}
F_j(z_2)H^\pm_i(z_1),
\end{eqnarray}
\begin{eqnarray}
~[E_i(z_1),F_j(z_2)]=\frac{\delta_{i,j}}{(q-q^{-1})z_1z_2}\left(
\delta\left(q^{-k}\frac{z_1}{z_2}\right)H_i^+(q^{-\frac{k}{2}}z_1)-
\delta\left(q^{k}\frac{z_1}{z_2}\right)H_i^-(q^{-\frac{k}{2}}z_2)\right).
\end{eqnarray}
\end{thm}
Now we have costructed level $k$ 
free field realization of Drinfeld current 
$E_i(z), F_i(z), H_i^\pm(z)$
for the elliptic algebra
$U_{q,p}(\widehat{sl_3})$.
This gives the first example of arbitrary-level free field
realization of higher-rank elliptic algebra.

\section*{Acknowledgement}~The author would like to thank
the organizing committee of the 27-th 
International Colloquium of the Group Theoretical Method
in Physics held at Yerevan, Armenia 2008.
The author would like to thank Prof.A.Kluemper for his kindness at Armenia.
This work is partly supported by the Grant-in Aid
for Young Scientist {\bf B}(18740092) from Japan Society for
the Promotion of Science.

\section*{Appendix}
In appendix we summarize the normal ordering
of the basic operators.

\begin{eqnarray}
:e^{\gamma^i(z_1)}:
{\cal B}_+^{* i}(z_2)&=&:
e^{\gamma^i(z_1)}
{\cal B}_+^{* i}(z_2):
\frac{(q^{r^*+1}z_2/z_1;p^*)_\infty}{(q^{r^*-1}z_2/z_1;p^*)_\infty},
\nonumber\\
:e^{\beta_1^i(z_1)}:
{\cal B}_+^{* i}(z_2)&=&
:e^{\beta_1^i(z_1)}
{\cal B}_+^{* i}(z_2):
\frac{(q^{r^*}z_2/z_1;p^*)_\infty}{
(q^{r^*+2}z_2/z_1;p^*)_\infty},
\nonumber\\
:e^{\beta_2^i(z_1)}:
{\cal B}_+^{* i}(z_2)&=&
:e^{\beta_2^i(z_1)}
{\cal B}_+^{* i}(z_2):
\frac{(q^{r^*}z_2/z_1;p^*)_\infty}{
(q^{r^*+2}z_2/z_1;p^*)_\infty},
\nonumber\\
:e^{\beta_3^i(z_1)}:
{\cal B}_+^{* i}(z_2)&=&:
e^{\beta_3^i(z_1)}
{\cal B}_+^{* i}(z_2):
\frac{(q^{r^*}z_2/z_1;p^*)_\infty}{
(q^{r^*-2}z_2/z_1;p^*)_\infty},
\nonumber\\
:e^{\beta_4^i(z_1)}:
{\cal B}_+^{* i}(z_2)&=&:
e^{\beta_4^i(z_1)}
{\cal B}_+^{* i}(z_2):
\frac{(q^{r^*}z_2/z_1;p^*)_\infty}{
(q^{r^*+2}z_2/z_1;p^*)_\infty},
\nonumber
\\
{\cal B}_-^i(z_1):e^{\gamma^i(z_2)}:&=&:
{\cal B}_-^i(z_1)e^{\gamma^i(z_2)}:
\frac{(q^{r+1}z_2/z_1;p)_\infty}{(q^{r-1}z_2/z_1;p)_\infty},
\nonumber
\\
{\cal B}_-^i(z_1):e^{\beta_1^i(z_2)}:&=&:
{\cal B}_-^i(z_1)e^{\beta_1^i(z_2)}
:\frac{(q^rz_2/z_1;p)_\infty}{
(q^{r+2}z_2/z_1;p)_\infty},
\nonumber
\\
{\cal B}_-^i(z_1):e^{\beta_2^i(z_2)}:&=&
:{\cal B}_-^i(z_1)e^{\beta_2^i(z_2)}:
\frac{(q^rz_2/z_1;p)_\infty}{
(q^{r+2}z_2/z_1;p)_\infty},
\nonumber
\\
{\cal B}_-^i(z_1):e^{\beta_3^i(z_1)}:&=&:
{\cal B}_-^i(z_1)e^{\beta_3^i(z_1)}:
\frac{(q^rz_2/z_1;p)_\infty}{
(q^{r-2}z_2/z_1;p)_\infty},
\nonumber
\\
{\cal B}_-^i(z_1):e^{\beta_4^i(z_2)}:&=&:
{\cal B}_-^i(z_1)e^{\beta_4^i(z_2)}
:
\frac{(q^rz_2/z_1;p)_\infty}{
(q^{r-2}z_2/z_1;p)_\infty},\nonumber
\end{eqnarray}
\begin{eqnarray}
e^{b_+^i(z_1)}{\cal B}_+^{* i}(z_2)&=&
:
e^{b_+^i(z_1)}{\cal B}_+^{* i}(z_2)
:\frac{
(q^{r^*}z_2/z_1;p^*)_\infty^2}{
(q^{r^*+2}z_2/z_1;p^*)_\infty 
(q^{r^*-2}z_2/z_1;p^*)_\infty},\nonumber
\\
{\cal B}_-^i(z_1) e^{b_-^i(z_2)}&=&:
{\cal B}_-^i(z_1) e^{b_-^i(z_2)}
:\frac{(q^rz_2/z_1;p)_\infty^2}{
(q^{r+2}z_2/z_1;p)_\infty 
(q^{r-2}z_2/z_1;p)_\infty},\nonumber
\\
e^{a_+^i(z_1)}{\cal A}^{*i}(z_2)&=&
:e^{a_+^i(z_1)}{\cal A}^{*i}(z_2):
\frac{
(q^{r^*+k+5}z_2/z_1;p^*)_\infty
(q^{r^*-k-5}z_2/z_1;p^*)_\infty}{
(q^{r^*+k+1}z_2/z_1;p^*)_\infty
(q^{r^*-k-1}z_2/z_1;p^*)_\infty},\nonumber\\
e^{a_+^1(z_1)}{\cal A}^{*2}(z_2)&=&:
e^{a_+^1(z_1)}{\cal A}^{*2}(z_2):
\frac{
(q^{r^*+k+2}z_2/z_1;p^*)_\infty
(q^{r^*-k-2}z_2/z_1;p^*)_\infty}{
(q^{r^*+k+4}z_2/z_1;p^*)_\infty
(q^{r^*-k-4}z_2/z_1;p^*)_\infty},\nonumber\\
e^{a_+^2(z_1)}{\cal A}^{*1}(z_2)&=&
:
e^{a_+^2(z_1)}{\cal A}^{*1}(z_2):
\frac{
(q^{r^*+k+2}z_2/z_1;p^*)_\infty
(q^{r^*-k-2}z_2/z_1;p^*)_\infty}{
(q^{r^*+k+4}z_2/z_1;p^*)_\infty
(q^{r^*-k-4}z_2/z_1;p^*)_\infty},\nonumber\\
{\cal A}^{i}(z_1)e^{a_-^i(z_2)}&=&:
{\cal A}^{i}(z_1)e^{a_-^i(z_2)}
:\frac{(q^{r+k+5}z_2/z_1;p)_\infty (q^{r-k-5}z_2/z_1;p)_\infty}{
(q^{r+k+1}z_2/z_1;p)_\infty (q^{r-k-1}z_2/z_1;p)_\infty},
\nonumber\\
{\cal A}^{1}(z_1)e^{a_-^2(z_2)}&=&:
{\cal A}^{1}(z_1)e^{a_-^2(z_2)}:
\frac{
(q^{r+k+2}z_2/z_1;p)_\infty
(q^{r-k-2}z_2/z_1;p)_\infty}{
(q^{r+k+4}z_2/z_1;p)_\infty
(q^{r-k-4}z_2/z_1;p)_\infty},\nonumber\\
{\cal A}^{2}(z_1)e^{a_-^1(z_2)}&=&
:{\cal A}^{2}(z_1)e^{a_-^1(z_2)}:
\frac{
(q^{r+k+2}z_2/z_1;p)_\infty
(q^{r-k-2}z_2/z_1;p)_\infty}{
(q^{r+k+4}z_2/z_1;p)_\infty
(q^{r-k-4}z_2/z_1;p)_\infty},\nonumber
\end{eqnarray}
\begin{eqnarray}
{\cal B}_-^i(z_1){\cal B}_{+}^{*i}(z_2)&=&:
{\cal B}_-^i(z_1){\cal B}_{+}^{*i}(z_2)
:\frac{(q^k z_2/z_1;q^{2k},p^*)_\infty^2}{
(q^{k+2}z_2/z_1;q^{2k},p^*)_\infty
(q^{k-2}z_2/z_1;q^{2k},p^*)_\infty
}\nonumber\\
&\times&\frac{
(q^{k+2}z_2/z_1;q^{2k},p)_\infty
(q^{k-2}z_2/z_1;q^{2k},p)_\infty}
{(q^k z_2/z_1;q^{2k},p)_\infty^2},
\nonumber\\
{\cal A}^i(z_1){\cal A}^{*i}(z_2)&=&:
{\cal A}^i(z_1){\cal A}^{*i}(z_2):
\frac{(q^{2k+5}z_2/z_1;q^{2k},p^*)_\infty
(q^{-5}z_2/z_1;q^{2k},p^*)_\infty
}{
(q^{2k+1}z_2/z_1;q^{2k},p^*)_\infty
(q^{-1}z_2/z_1;q^{2k},p^*)_\infty
}\nonumber\\
&\times&
\frac{(q^{2k+1}z_2/z_1;q^{2k},p)_\infty
(q^{-1}z_2/z_1;q^{2k},p)_\infty
}{
(q^{2k+5}z_2/z_1;q^{2k},p)_\infty
(q^{-5}z_2/z_1;q^{2k},p)_\infty},\nonumber
\\
{\cal A}^1(z_1){\cal A}^{*2}(z_2)&=&:
{\cal A}^1(z_1){\cal A}^{*2}(z_2):
\frac{(q^{2k+2}z_2/z_1;q^{2k},p^*)_\infty
(q^{-2}z_2/z_1;q^{2k},p^*)_\infty
}{
(q^{2k+4}z_2/z_1;q^{2k},p^*)_\infty
(q^{-4}z_2/z_1;q^{2k},p^*)_\infty
}\nonumber\\
&\times&
\frac{(q^{2k+4}z_2/z_1;q^{2k},p)_\infty
(q^{-4}z_2/z_1;q^{2k},p)_\infty
}{
(q^{2k+2}z_2/z_1;q^{2k},p)_\infty
(q^{-2}z_2/z_1;q^{2k},p)_\infty},\nonumber
\\
{\cal A}^2(z_1){\cal A}^{*1}(z_2)&=&:
{\cal A}^2(z_1){\cal A}^{*1}(z_2):
\frac{(q^{2k+2}z_2/z_1;q^{2k},p^*)_\infty
(q^{-2}z_2/z_1;q^{2k},p^*)_\infty
}{
(q^{2k+4}z_2/z_1;q^{2k},p^*)_\infty
(q^{-4}z_2/z_1;q^{2k},p^*)_\infty
}\nonumber\\
&\times&
\frac{(q^{2k+4}z_2/z_1;q^{2k},p)_\infty
(q^{-4}z_2/z_1;q^{2k},p)_\infty
}{
(q^{2k+2}z_2/z_1;q^{2k},p)_\infty
(q^{-2}z_2/z_1;q^{2k},p)_\infty}.\nonumber
\end{eqnarray}
Here we have used the notation 
$$(z;p_1,p_2)_\infty=\prod_{n_1,n_2=0}^\infty
(1-p_1^{n_1}p_2^{n_2}z).$$

\end{document}